\documentclass[english]{IEEEtran}
\usepackage[T1]{fontenc}
\usepackage{amsmath}
\usepackage{stmaryrd}
\usepackage{graphicx}
\usepackage{cite}

\makeatletter

\providecommand{\tabularnewline}{\\}

\makeatother

\usepackage{babel}
\begin{document}
\title{Emotion Recognition using Machine Learning and ECG signals}

\author{Bo Sun and Zihuai Lin, \emph{Senior Member}, 

\thanks{Bo Sun and Zihuai Lin are with the School of Electrical and Information Engineering,
The University of Sydney, Australia (e-mail: zihuai.lin@sydney.edu.au).}
}
\maketitle
\begin{abstract}
Various emotions can produce variations in electrocardiograph (ECG) signals, distinct emotions can be distinguished by different changes in ECG signals. This study is about emotion recognition using ECG signals. Data for four emotions, namely happy, exciting, calm, and tense, is gathered. The raw data is then de-noised with a finite impulse filter. We use the Discrete Cosine Transform (DCT) to extract characteristics from the obtained data to increase the accuracy of emotion recognition. The classifiers Support Vector Machine (SVM), Random Forest, and K-NN are explored. To find the optimal parameters for the SVM classifier, the Particle Swarm Optimization (PSO) technique is used. The results of the comparison of these classification methods demonstrate that the SVM approach has a greater accuracy in emotion recognition, which may be applied in practice.
\end{abstract}

keywords-emotion recognition; ECG signals; DCT; SVM; PSO; Random
Forest; K-NN

\section{Introduction}

In recent years, emotion recognition (detection) gains lots of attention.
However, most emotion detection methods are based on behavior
such as speech detection and face recognition \cite{speech,d(facial)}.
The major drawback for behavior based emotion recognition is
uncertainty. Because the behavior induced by emotion can be suppressed.
For example, facial expressions can be used to judge emotion, but
no one guarantees that related cues will definitely be expressed, whether or
not people are experiencing an emotion. An alternative to separating
emotion is using physiological signals. The common physiological signals
include the electroencephalogram (EEG) \cite{EEGxucun}, electromyogram (EMG) and electrocardiogram (ECG).
Traditionally, these types of biological signals are used for clinical
diagnosis. For example, the ECG signal is used to detect and analyze
diseases such as heart diseases. There are evidences showing that physiological
signals are sensitive to emotion states, which indicates that the
signals may convey emotional information \cite{emotion information,emotion information1}. 

Compared
with previous methods of emotion recognition, there are two benefits
to using physiological signals on emotion detection. One is that physiological
signals are involuntary reactions, such reactions are hard to mask.
Furthermore, physiological signals can be recorded continuously by
sensors attached to body. This is not like the case of speech recognition.
The information only can be recorded while people are speaking. 

This
paper focuses on the emotion recognition using ECG signals. An ECG device records the electrical changes caused by activities of heart, which is collected by electrodes over the skin in a period of time.  
According to \cite{compare face and ECG,face}, currently, the accuracy of emotion
recognition based on ECG signals is lower than above mentioned methods
such as face recognition. The highest accuracy of emotion detection
was around 80 percentage \cite{compare face and ECG}. However, the face recognition accuracy can
reach to 90\% \cite{ECG}. In this paper, we investigate different machine learning
methods to improve the performance of emotional classifications
based on ECG signals.

Data collection is one of the most important steps for emotion detection.
The definition of different emotions must be explicit in this phase.
If the definition is not clear, confusion may occur among different
emotions in the classification phase and the classification performance will be influenced negatively. In the data collection stage, we collect the ECG signals for four emotions: happy, exciting, calm and tense, respectively. The ECG signals are collected by using a low-cost wearable ECG patch, called iRealcare \cite{IREALCARE1, IREALCARE2, IREALCARE3,IREALCARE4,IREALCARE5}. The collected signals are pre-processed by a finite impulse filter to remove noises from raw ECG signals. Next we separate the processed data into two datasets, one set is used to build a training
set and the other set is used as a test set. Moreover, to accurately recognize different
emotions, the main features of different emotions must be extracted by the feature extraction from the training set and the test set. In addition, the feature extraction can delete useless information to reduce data redundancy. In our experiments, we employ the Discrete Cosine Transform (DCT) to extract features \cite{DCT}.
In the classification stage, three classifiers are used to analyze
emotional data. First, the support vector
machine is selected as the classifier. 

To improve the performance
of classification, the Radial Basis Function (RBF) kernel is used
to improve the data dimension. In consideration of the importance of kernel
parameters, the Particle Swarm Optimization (PSO) algorithm is applied
to search the optimal parameters. Second, the support
vector machine classifier is replaced with Random Forest classifier. The SVM
classifier originates from separating two categories. Thus, for multiple categories, the performance may be poor. Compared with support vector machine, the Random Forest is good at disposing
the multi-class classification. Third, the K-NN classifier
is selected as the classifier. The training complexity of K-NN is lower, compared to above two classifiers. Finally, by analyzing the results of different classifiers,
we select one best classifier for emotion detection based on ECG signals. 

The paper is organized as follows: Section II introduces the ECG data collection, data pre-processing and the selection of training set and test
set. The method of feature extraction used in ECG signals are presented
 in Section III. In Section IV, we introduce
classifiers used for analyzing ECG signals. Section V compares
performance of different classifiers and presents a best method
for emotion recognition based on ECG signals. Section VI draws the conclusion of this work. 

\section{Data collection and pre-processing }

An ECG sensor from iRealcare \cite{IREALCARE1,IREALCARE2,IREALCARE3,IREALCARE4,IREALCARE5} was used for collecting
ECG signals of different emotions from different people. The data collected by the iRealcare ECG sensor can be transmitted to a smart phone APP via Bluetooth Low Energy (BLE) and then to a cloud. From the cloud, we can acquire the raw ECG signals.
In the experiment, ECG signals were recorded from four emotions including
calm, exciting, happy and tense. Except calm, each emotion is generated based on
external environmental stimulus. The calm emotion describes the normal state,
so the ECG signal was recorded without any external stimulus. The
ECG signals for exciting, happy and tense emotions were recorded when the testing people do exercises, watch comedies and watch thriller movies, respectively. Generally, a person's emotion changes quickly, so the record duration should not be too long. The longer the duration is, the more useless information will be added into ECG signals.
Thus, the record time is set to 5 minutes for each emotion. The sampling rate for the ECG signal is 128 Hz. Taking into account differences among different people, we selected 5 participants. There are 5 data groups for each emotion and total 20 data groups for four
emotions. 

Normally, human ECG signal is non-linear and low signal to noise. The frequency range of ECG signals is from 0.05HZ to 100HZ and the dynamic range is below 4mv \cite{all noise}. Thus, the collected ECG signals are susceptible to be disturbed by external factors such as interference. To acquire ECG signals with lower interference, we need to conduct the pre-processing for the raw ECG signals. During the collection and transmission, the ECG signals are affected by different noises. Two types of interference are mainly involved in ECG signals and they are low frequency noises and high frequency noise. The low frequency noises include baseline drift. The high frequency noises include Electromyography (EMG) noise, power line interference and channel noises. The baseline drift is a low frequency noise in ECG signals and it causes by body movement and breathing. Baseline drift can make the entire ECG signal shift down or up at the axis. The frequency of baseline drift is greater than 1HZ and the influence of baseline drift is on the analysis of peak. Power-line interference can be caused by the electromagnetic field of nearby facilities and  electromagnetic interference of power line. Since the iRealcare sensor used BLE instead of cables, the power-line interference only can be generated by the electromagnetic field of nearby machines. 

EMG noise is generated by electric activities of muscle. The noise is usually in subjects with disable persons or people who have uncontrollable tremor diseases. In addition, there is electrode contact noise in ECG signals. Electrode contact noise is caused by skin stretching. Because the impedance of skin will change when the skin is stretched. Their frequency is from 1 to 10 HZ \cite{noise}. We used the finite impulse filter(FIR) to filter the noise. FIR filter is a reliable and simple filter. At the same time, FIR filter is a linear filter and the output will not be distorted \cite{FIR}. In design of FIR filter, the window method is used. The common window methods include Hamming window, 
Rectangular window, Hanning window, and Blackman window. These different windows are used to design the low pass filter and high pass filter with cut-off frequencies in FIR filter. The frequencies above the cut-off frequency is blocked and others can pass in FIR filter. In our FIR filter, the cut-off frequencies are set to 3HZ and 100Hz respectively. After de-noising collected data, currently, we have the ECG data of four emotions and each emotion has five groups. Total number of ECG data is twenty groups. Then we selected four groups of each emotion as training group and the rest one group is used as test group. The segment of ECG signals from training set and test set were displayed respectively in Figs 1 and 2.

\begin{figure}
\includegraphics[width=8cm]{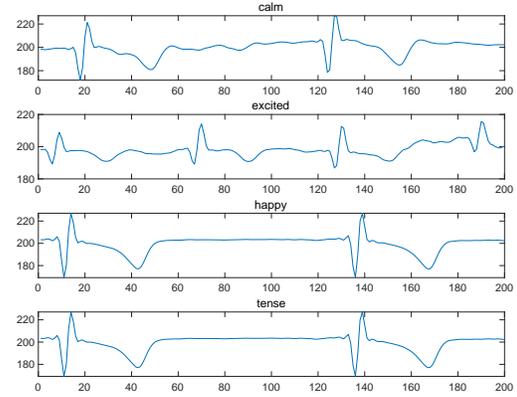}\caption{The ECG segments (training set) from four emotions}

\end{figure}
\begin{figure}

\includegraphics[width=8cm]{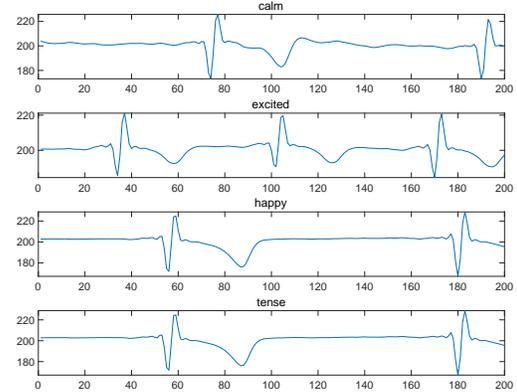}\caption{The ECG segments (test set) from four emotions}

\end{figure}

\section{Feature Extraction}

The feature extraction is an important part in machine learning. The
purpose of feature extraction is to extract the main information from
data further compressing data to improve classification performances.
For an ECG signal, there are three main components: P wave, T wave
and QRS, which represent three phases of the ECG signal. P wave
is the contraction pulse in atrial systole. QRS represents the depolarization
of ventricles. T wave is re-polarization of ventricles \cite{ECG characteristics}.
However, there are some valueless features in the
main components and the function of feature extraction is to remove
these features. There are lots of feature extraction methods in
different domains. In this paper, we use frequency domain discrete cosine transform (DCT) methods 
to extract main information of ECG signals. 

\subsection*{A. Discrete cosine transform}

The discrete cosine transform expresses a finite sequence as a sum of cosine functions at different frequencies. The function of discrete
cosine transform converts a real signal into a signal in frequency
domain. The discrete cosine transform is a Fourier transform without the
conjugate part. The formula is as follows:
\begin{equation}
y(k)=w(k)\mathop{{\displaystyle \sum_{n=1}^{N}x(n)cos[\frac{\pi}{2N}(2n-1)(k-1)],} k=1,...,N}    
\end{equation}
where
\begin{eqnarray}
 w(k)=\protect\begin{cases}
\frac{1}{\sqrt{N}} & ,k=1\protect\\
\sqrt{\frac{2}{N}} & ,2\protect\leq k\protect\leq N
\protect\end{cases}
\end{eqnarray}
and $N$ is the total number of elements \cite{dct}.

The ECG signals are made up by a large amount of sample points. After
DCT, it was found that each segment of ECG signals was
converted to the ECG signals at frequency domain and the sample points
of ECG signals were rearranged in a decreasing order. 

After the feature extraction, the ECG segments from the training set and test set are shown in Figs. 3 and 4, respectively. Comparing with original ECG signals, the ECG signals experienced feature extraction are easier to observe the differences between distinct emotions. In the
feature extraction, there is an important parameter need to be considered. It is the number of extracted feature values and
the best value can be determined by comparing classification performances
at different values. The number of extracted feature in Figs. 3 and
4 is 75.

\begin{figure}
\includegraphics[width=8cm]{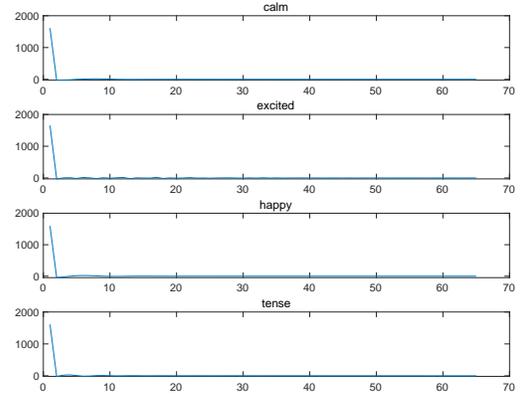}\caption{The ECG segments of training set (Feature extraction)}

\end{figure}
\begin{figure}
\includegraphics[width=8cm]{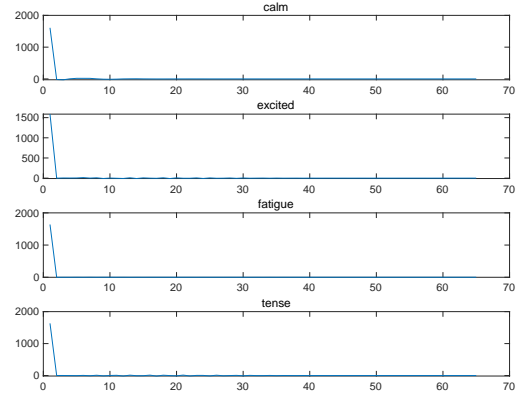}\caption{The ECG segments from test set (Feature extraction)}

\end{figure}

\section{The Classifier Selection}

In the selection of classifiers, Support Vector Machine (SVM), Random
Forest and K Nearest Neighbor (K-NN) were used to classify the ECG
data. For the SVM classifier, it is difficult for the data from four emotions to be discriminated in low dimension due to the nonlinear characteristic of ECG data. Thus, Radial Basis function (RBF) was used. The RBF can map the data to higher dimension to separate objects. The parameters C and $\gamma$ in RBF kernel
play a key role on the classification and they determine the rules of classification. Thus Particle
Swarm Optimization (PSO) algorithm is used to optimize parameters
C and $\gamma$. Then we used Random Forest classifier and K-NN classifier
to attempt classification. By comparing the classification performance
of different classifiers, the best classification method was proposed. 

\subsection*{A. Support Vector Machine (SVM)}

The SVM is a supervised learning method, which is used for classification
and regression. The basic concept is to find an optimal hyper-plane
that separates data into different classes. The hyper-plane is decided
by the support vectors which lies on the hyper-plane. The optimal
hyper-plane is defined as the separating hyper-plane with maximum margin (see
Fig. 1) \cite{svm}. 

\begin{figure}
\includegraphics[width=8cm]{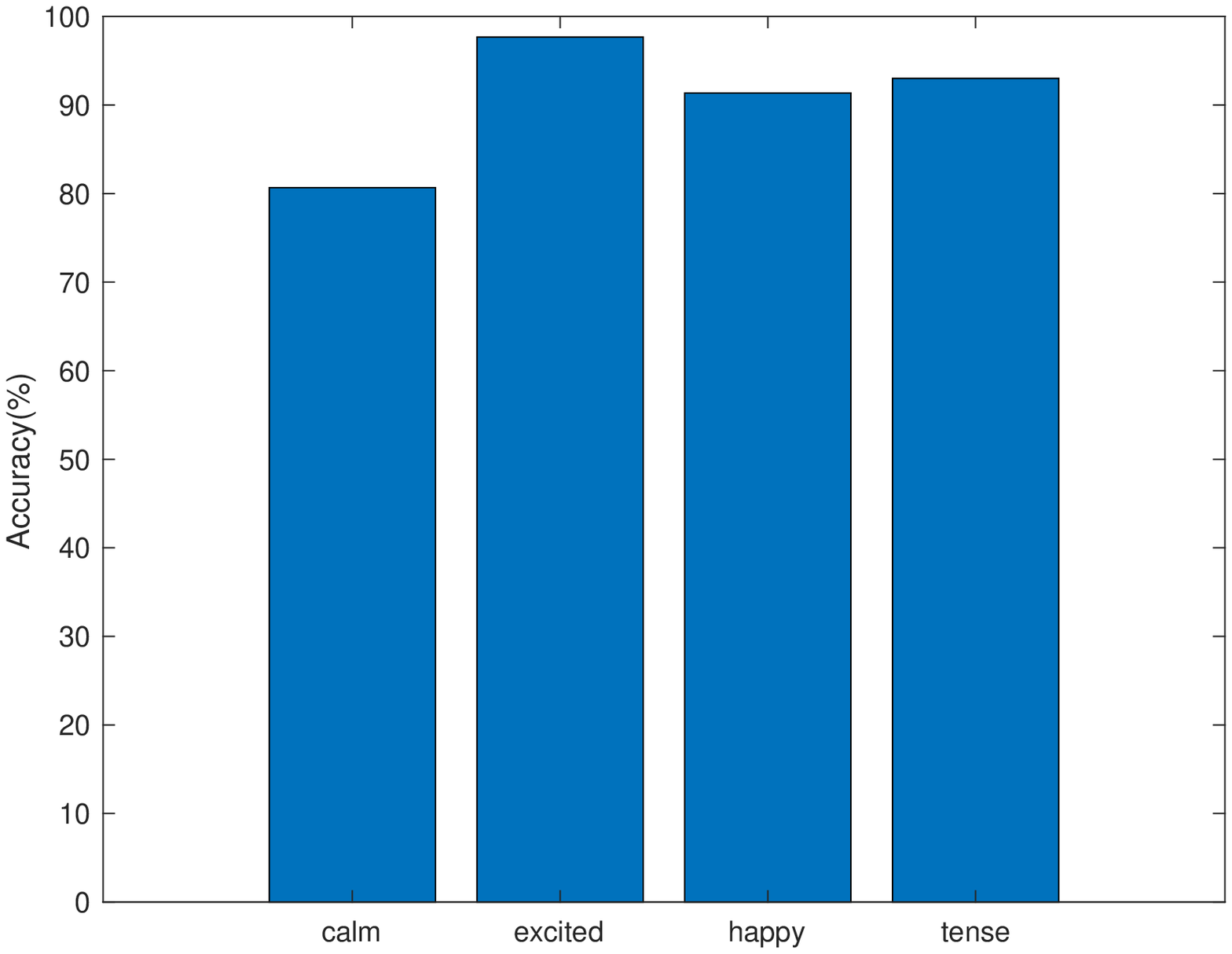}
\caption{SVM classification}
\end{figure}
In SVM, known objects are used as training sets. A training set is made
up by feature values and feature labels, which is used to build the
classification model. For example, there are N training data $(x_{i},y_{i}),i=1,...,N.$
$x_{i}\in R^{n}$ and $y_{i}\in\left\{ -1,1\right\}^{N} $\cite{PSO_svm(1)}.
If data are linearly separable, there exists a weight vector w and a scalar b which satisfy the following inequalities.
\begin{eqnarray}
    & w*x_{i}+b\geq 1,   & y_{i}=1 \\ \nonumber
    & w*x_{i}+b\leq -1,  & y_{i}=-1
\end{eqnarray}



The optimal hyper-plane $(w_{0}, b_{0})$ is described as follow:
\begin{equation}
    w_{0}*x+b_{0}=0
\end{equation}
where $x$ is an input pattern.
The decision function is
\begin{equation}
    f(x)=sign((w\cdot x)+b)
\end{equation}

According to \cite{vapnik}, the support vector machine(SVM) follows the solution of eq. (\ref{eq.svm_min}) under the constraints of eqs. (\ref{eq.svm_min1}) and (\ref{eq.svm_min2}) in optimization problem.
\begin{eqnarray}
    Minimize & \frac{1}{2}w^{2}+C\sum_{i=1}^{N}\xi_{i} \label{eq.svm_min} \\ 
    s.t.& y_{i}(wx_{i}+b)\geq 1-\xi_{i}  \label{eq.svm_min1} \\
        & \xi_{i}\geq 0, i=1,...,N \label{eq.svm_min2}
\end{eqnarray}
where $\sum_{i=1}^{N}\xi_{i}$ is the total number of errors and $C$ is a penalty coefficient.

In practice, most problems are non-linear separable. It is difficult for SVM classifier to find  the optimal hyper-planes which separate different classes in original dimensions. Thus, to complete classification, the training
data set needs to be mapped into a high dimension space to find  separating
hyper-planes. By the use of a kernel function, the data dimension can be improved. There are several kernel functions. In our experiments, Radial basis function (RBF)
kernel function was used. The formula of RBF function is shown in
eq. (\ref{RBF}). 
\begin{equation} \label{RBF}
   k(x_{i},x_{j})=exp(-\gamma*\parallel x_{i}-x_{j}\parallel^{2}.
\end{equation}

Then we can get the decision function of RBF kernels as shown in eq. (\ref{RBF1}). 
\begin{equation} \label{RBF1}
   f(x)=sign((k(x,y)\cdot x)+b).
\end{equation}

$C$ and $\gamma$ are two important parameters of RBF kernel function, which
have an influence on separating performance. $C$ is called the penalty coefficient
and it is the tolerability of error. The bigger $C$ is, the lower error
tolerability will be, which will lead to over-classification and poor 
generation ability. If $C$ is smaller, the classification
error will increase. $\gamma$ is a spacial parameter which controls data distribution in new feature
space \cite{pso-svm(2)}. The value of $\gamma$ is affected by $\sigma$.
The relationship between $\gamma$ and $\sigma$ is as follows:
\begin{eqnarray} \label{rel_gamma}
k(x,y)&= & exp(-\parallel x-y\parallel^{2}/(2\sigma^{2})  \nonumber \\
 &=&exp(-\gamma\cdot\parallel x-y\parallel^{2})  \nonumber \\
&\Mapsto & \gamma=\frac{1}{2\cdot\sigma^{2}}
\end{eqnarray}

From eq. (\ref{rel_gamma}), we can see that as
$\sigma$ becomes smaller, $\gamma$ gets bigger. The
small $\sigma$ makes Gaussian distribution narrow and high.
In this way, there is a better classification performance for known
samples. However, for unknown data, the classification performance
is poor. The Gaussian distribution will be wider, while $\sigma$
is larger. For a larger $\sigma$, the model built by training set
is not accurate and it will influence test results. 
As mention above, the values of $C$ and $\gamma$ are vital and it will indicate SVM classifier how to separate data. Thus searching best parameters is necessary. The particle swarm
optimization (PSO) algorithm was used to search the best values of  $C$ and $\gamma$  \cite{pso}.
In the PSO algorithm, the optimization problem can be solved by 
searching the particle position. The flow chart of the PSO algorithm
is shown in Fig. 6. 

\begin{figure}
\includegraphics[width=10cm]{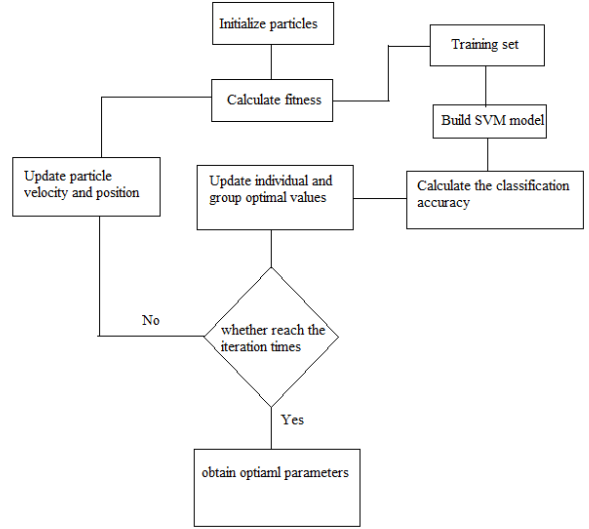}
\caption{The flow chart of PSO algorithm}
\end{figure}
At first, particles are initialized in a feasible space. Each particle
represents a possible optimal solution. The characteristics of a particle
is described by its velocity, position and fitness. The velocity is a
preset value. The location of each particle is updated by the individual
and group extreme value. The fitness value is calculated when the position
of each particle is updated. The position of the individual and group
extreme value is updated by comparing the new fitness with fitness extremum.
The updated formulas are as follows:
\begin{eqnarray}
v_{i}(t+1) &=&v_{i}(t)+c_{1}r_{1}(P_{id}-x_{i}(t))+c_{2}r_{2}(P_{gd}-x_{i}(t))  \nonumber \\
x_{i}(t+1)&=&x_{i}(t)+v_{i}(t+1)
\end{eqnarray}
where $x_{i}$ is the location of the particle; $v_{i}$ denotes the velocity
of the particle; $c_{1}$ and $c_{2}$ are learning factors; $P_{id}$ is
the location of the optimal individual value and $P_{gd}$ is the location
of the optimal group value. $r_{1}$ and $r_{2}$are random values between
0 and 1. 

\subsection*{B. Random Forest}

\begin{figure}
\includegraphics[width=10cm]{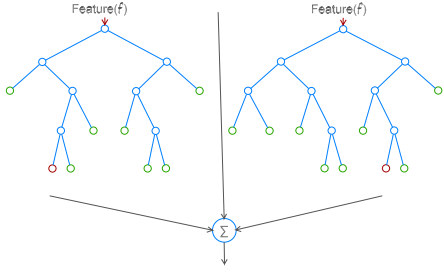}
\caption{The structure of Random forest}
\end{figure}

The structure of Random forest was shown in Figure 7. From Figure 7, it could be seen that Random forest can be seen as a set of decision tree. By building multiple decision trees, merge their results together to get a reliable result. In machine learning,
decision tree is a model, which represents the mapping between the
object and the attribute of object \cite{random}. The nodes in trees
are the analyzing conditions and the leaves of trees are the prediction
results. Compared with other supervised learning algorithms, Random forest adds randomness to the analysis. It randomly selects the features from a feature set to build decision trees and finally average the results. Thus, the generalization ability of random forest is stronger than other classifiers. The random forest is an ensemble learning over bagging. Ensemble
learning deals with prediction problem by combining several models \cite{random forest(1)}.
There are many decision trees in random forest and it is an ensemble
of decision trees. For an ensemble classifier $c_{1},c_{2},...,c_{n},$ the
training set is selected randomly from vectors $X$ and $Y$. The margin
function is described by
\begin{equation}
    M(X,Y)=I(C_{n}(X)=Y)-\max I(C_{n}(X)=i), {i\neq Y}\nonumber \\
\end{equation}
%
where $I$ is an indicator function. The margin function calculates
the confidence of voting. The larger the margin is, the better classification
performance will be. According to the margin function, we define the generalization
error as follows:
\begin{equation}
GE=P_{X,Y}(M(X,Y) < 0)
\end{equation}
In this paper, the random forest algorithm is inspired from the bagging
and the subsets. The random forest was built by bagged decision trees.
At first, Bootstrap samples from all samples are selected. Then features
of samples are selected. Next, decision trees are generated by repeating above two steps.
Then data will be input to all decision trees. The decision trees
are the classifiers and each decision tree has a result of classification. Finally,
according to the results of all classifiers, the best classification result is determined by voting. 

\subsection*{C. K Nearest Neighbor(K-NN)}

K-NN is the simplest classification non-parametric algorithm, which
classifies unclassified samples by measuring the distances between
training samples and test samples. From eq. (13), the rule
of nearest neighbor is defined as 

\begin{equation}
\min d(x_{i},x)=d(x_{n^{'}},x),i=1,2,...n 
\end{equation}
where $x_{i}$ is the $ith$ measurement, $x_{n^{'}}\in(x_{1},x_{2},...,x_{n})$
and $x$ is the objects which will be classified. Four different functions
can be used to calculate the distance including cosine, Euclidean,
Minkowsky and Chi square \cite{K calculation}. The Euclidean distance function is the most
widely used to calculate the distance metric of K-NN. The function
of Euclidean distance in the space can be computed by
\begin{eqnarray}
d(X,Y)&=&\sqrt{(x_{1}-y_{1})^{2}+(x_{2}-y_{2})^{2}+...+(x_{n}-y_{n})^{2}} \nonumber \\
&=&\sqrt{\sum_{i=1}^{n}(x_{i}-y_{i})^{2}},
\end{eqnarray}
where vectors $X=\left(x_{1,}...,x_{n}\right)$, $Y=\left(y_{1},...,y_{n}\right)$
and $n$ is the data dimension. 

In K-NN algorithm, the value of $K$ plays an important role in the performance
of classification as illustrated in Fig. 8, which shows that the
unclassified sample is assigned to different classes based on
different $K$ values. 

\begin{figure}
\includegraphics[width=10cm,height=8cm]{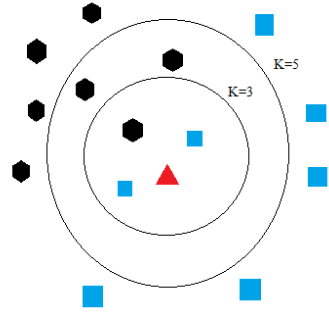}
\caption{an example K-NN model}
\end{figure}
When $K$ is equal to 3, the red triangle is assigned to the class
of blue square. If the value of $K$ is 5, the test sample belongs
to the black hexagon class. From Fig. 8, we can see that the
$K$ value has an influence on the result of classification. Thus, the selection
of $K$ is important. When a larger $K$ is used, the effect from nearest
neighbor decreases. The distance of the whole system needs to be
calculated, so the classification has to take much more time. If $K$ is very small, other
classes may affect the classification of test samples. Usually, we select a suitable $K$ by comparing performance of different $K$ values.

\section{Classification results}

The classification was implemented in Matlab. The aim of simulation
is to find a suitable method to recognize different emotions. In the simulation,
the total number of training data and test data was 4000 and 1200, respectively,
which were randomly selected from training groups and test groups.
Three classifiers were used to conduct the classification for four
emotion data, they are SVM, Random Forest and K-NN. Finally, by
comparing their results, we proposed the best method to recognize
emotion based on ECG signals. 

\subsection{SVM}

In our experiment, we first select SVM as the classifier.
The SVM classification is implemented by the LIBSVM \cite{LIBSVM}. After
the PSO algorithm, the optimal parameters $C$ and $\gamma$
are determined as 100.3 and 0.016, respectively. The features from the training set are used for building the model. The test set features are used for predicting. Then we use the confusion matrix to analyze the classification accuracy
by comparing the test label and prediction label. To select a suitable
number of features, the number of features are tested from 20 to
80 at an interval of 5. To get a more realistic result, for each number of features, we conduct classification ten times and calculate average recognition rate of four emotions. The results are shown in Fig. 9, from which we can see that initially there is an increasing
trend on average recognition rate as the number of features increases.
\begin{figure}
\includegraphics[width=8cm]{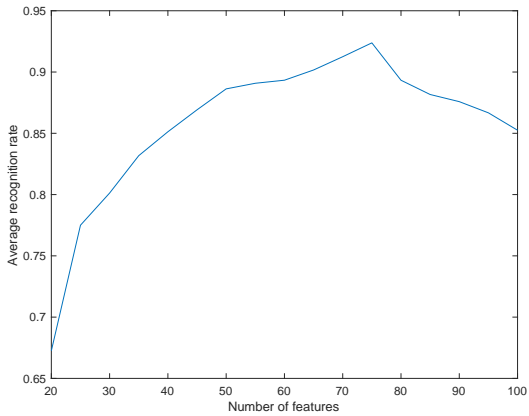}
\caption{The recognition rate versus the number of features}
\end{figure}
After reaching the peak, the recognition rate starts to fall. The
classification performance is the best, when the number of extracted
features is set to 75. 

Fig. 10 presents the classification results of four emotions. 
\begin{figure}
\includegraphics[width=8cm]{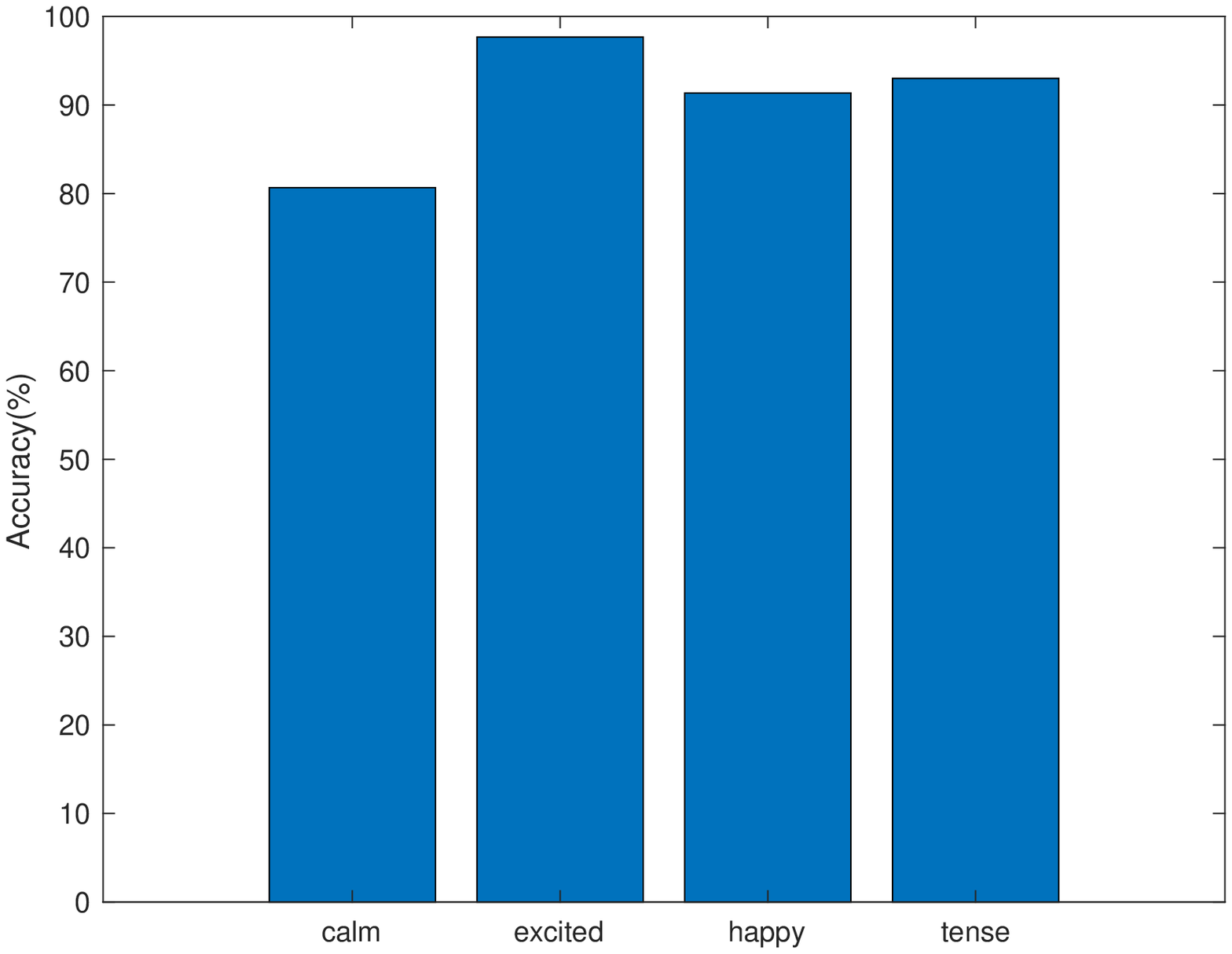}
\caption{The classification of emotion recognition by SVM}
\end{figure}
From Fig. 10, we can be see that the detection rate of exciting is the highest and it is over 90\%. The accuracy of happy and  tense approach 90\%. By comparing with these three emotions, the recognition rate of calm  is lowest and the accuracy is around 80\%.
Table I shows the average recognition rate of four emotions simulated
by PSO-SVM, when the analysis is conducted ten times. From Table I,
we can see that the average accuracy rate of three emotions (happy, exciting
and tense) reaches $90\%$, while the average accuracy of calm is around 80\%. 

\begin{table}\label{tab1}
\caption{The classification results simulated by PSO-SVM}
\begin{tabular}{|c|c|c|c|c|}
\hline 
Emotion & Happy & exciting & Calm & Tense\tabularnewline
\hline 
\hline 
Highest recognition rate & 93.7\% & 99.3\% & 87\% & 93\%\tabularnewline
\hline 
Lowest recognition rate & 90.7\% & 97\% & 77.3\% & 86.7\%\tabularnewline
\hline 
Average recognition rate & 92.51\% & 98.3\% & 80.93\% & 90.01\%\tabularnewline
\hline 
\end{tabular}

\end{table}

\subsection{Random Forest}

In the later analysis, the Random Forest classifier is used to for classification. From the result of SVM, we found that the classification
performance would be best as the number of features is 75. Thus,
the number of features are also set to 75. In random forest, there are
two methods in classification. One is Bagging, the other is boost.
In our experiments, the Bagging is used. Moreover, the number
of learning cycles needs to be determined in Random Forest. The best
number of learning cycle would be estimated by calculating the generalization
error. At first, the number of learning cycles is set to 200 and
calculate the loss error.  
\begin{figure}
\includegraphics[width=8cm]{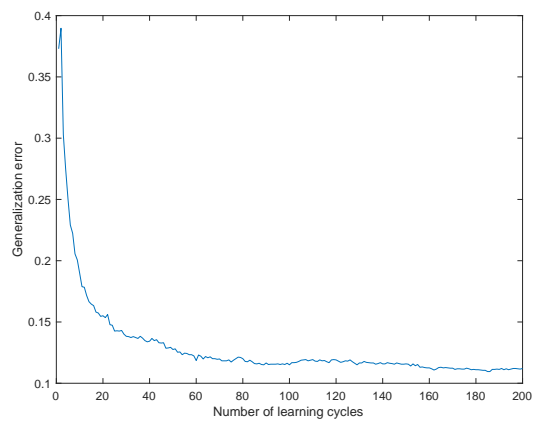}\caption{The Generalization error versus the number of learning cycles}
\end{figure}
The results can be seen in Fig. 11, from which, we can see that the loss decreased dramatically,
when the number of learners increased from 0 to 20. Then, it started
to slow down. Finally, the error would remain almost the same. From Fig. 11, we can estimate roughly the range of the best
number of learners. The number of learning cycles was started from
30 to 100 and each step was 10. For each learning cycle, we run the  classification algorithm for ten times then calculate the average recognition rate.
The results were shown in Fig. 12. 

\begin{figure}
\includegraphics[width=8cm]{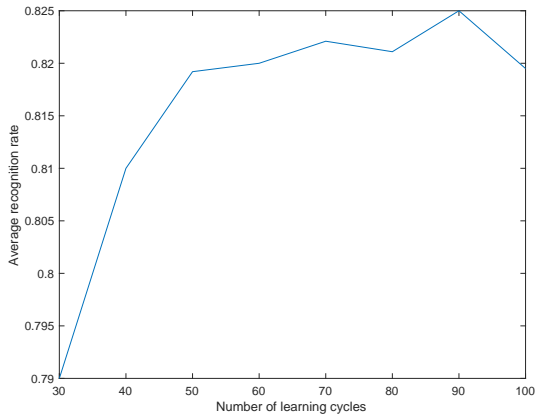}
\caption{The average recognition rate versus the number of learning cycles}

\end{figure}
From Fig. 12, we can see that there is not a significant
increase on recognition rate as the number of learner increases. There
is only $3\%$ increase on the average recognition rate, when the number
of learning cycles increases from 30 to 90. The best performance of
classification is achieved when the number of features and learners are set to 75 and 90, respectively.

\begin{figure}
\includegraphics[width=8cm]{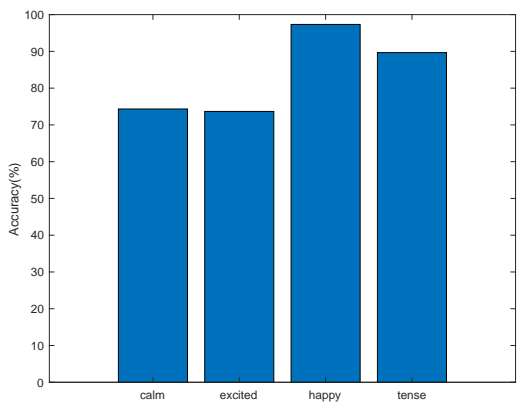}
\caption{The classification of emotion by Random forest}
\end{figure}
The recognition rate is shown in Fig. 13. We can see that the accuracy of happy is the highest and it closes to $100\%$. The recognition rate of tense approaches $90\%$. The recognition rates of calm and exciting are around $70\%$. 
Table II recorded the highest detection rate, lowest detection
rate and average recognition rate for four different emotions, when
running times is set to ten. From Table II, we can see that the average recognition
rate of happy is over 90\% and the accuracy rate of tense is about $90\%$. However, the average detection rates of calm and exciting
are relative low (under $80\%$). Especially, the recognition
rate of exciting is only $68.46\%$. 

\begin{table}
\caption{The classification results simulated by Random Forest}
\begin{tabular}{|c|c|c|c|c|}
\hline 
Emotion & Happy & exciting & Calm & Tense\tabularnewline
\hline 
\hline 
Highest recognition rate & 95\% & 780\% & 78.7\% & 90\%\tabularnewline
\hline 
Lower recognition rate & 912\% & 58.3\% & 74\% & 86.3\%\tabularnewline
\hline 
Average recognition rate & 94.56\% & 68.46\% & 75.7\% & 88.07\%\tabularnewline
\hline 
\end{tabular}

\end{table}

\subsection{K-NN}

Finally the classification is attempted by the K-NN classifier. In
K- NN algorithm , the value of $K$ needs to be determined first. For the
selection of $K$, the method is similar to the method used in Random
Forest. The classification loss is used to measure the performance
of K-NN and K is selected from 1 to 10. The number of extracted features
is still set to 75. The results are shown in Fig. 14. 

\begin{figure}
\includegraphics[width=8cm]{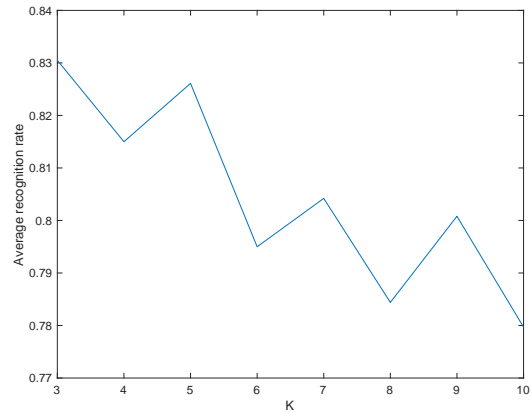}
\caption{The classification accuracy versus K }

\end{figure}
From Fig. 14, we can see that the classification accuracy is the
highest when $K=3$. After determining the value of $K$, we used
it to analyze the model. The test results are shown in Fig. 15. 

\begin{figure}
\includegraphics[width=8cm]{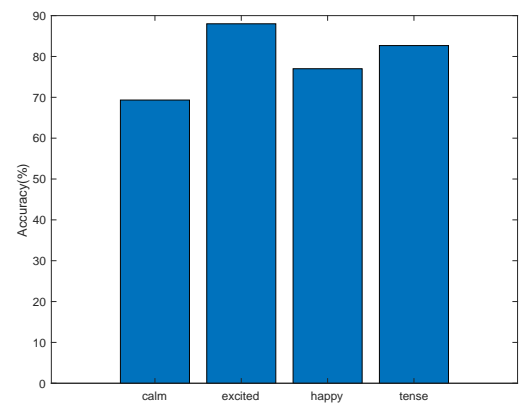}
\caption{The classification of emotion by K-NN}

\end{figure}
From Fig. 15, it can be seen the classification accuracy of
calm and happy are low, both of them are around $70\%$. The separating
accuracy of tense is over $80\%$. The recognition rate of exciting approaches
$90\%$. Table III records the recognition rates of four different
emotions, when the K-NN analysis runs ten times. Table III shows the high recognition rates for both exciting and tense. They are more than $85\%$. For the emotion of exciting, the average detection rate is around $80\%$. The lowest accuracy of emotion is for the calm which is under $80\%$. 

\begin{table}\label{tab:3}
\caption{The classification results simulated by K-NN}
\begin{tabular}{|c|c|c|c|c|}
\hline 
Emotion  & Happy & exciting & Calm & Tense\tabularnewline
\hline 
\hline 
Highest recognition rate & 88\% & 89.3\% & 80\% & 89.3\%\tabularnewline
\hline 
Lowest recognition rate & 83.7\% & 77\% & 72\% & 85.3\%\tabularnewline
\hline 
Average recognition rate & 85.9\% & 81.04\% & 76.03\% & 86.96\%\tabularnewline
\hline 
\end{tabular}

\end{table}

\subsection{Comparison and analysis}

We compare the performance of three classifiers, the results are shown in Fig. 16.


\begin{figure}
\includegraphics[width=8cm]{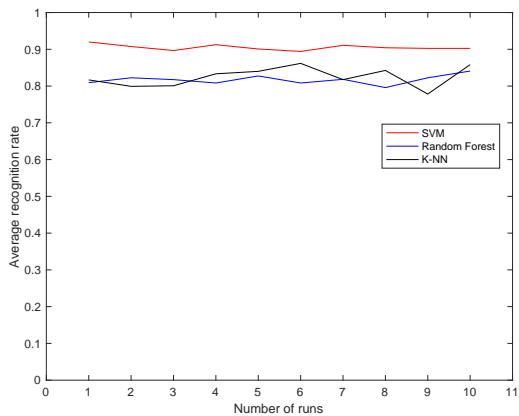}
\caption{The performance of the three classifiers }
\end{figure}

\begin{table}\label{tab:4}
\caption{The classification results completed by PSO-SVM and Random Forest
with different number of feature extraction}

\begin{tabular}{|c|c|c|c|}
\hline 
Classifier & PSO-SVM & Random forest & K-NN\tabularnewline
\hline 
\hline 
Optimal number of features & 75 & 75 & 75\tabularnewline
\hline 
Average recognition rate & 90.51667\% & 81.707\% & 82.483\% \\
(feature=best) & & & \tabularnewline
\hline 
\end{tabular}

\end{table}
It can be seen that performance of the PSO-SVM classifier is the best and the recognition rate is obviously higher than the recognition rates for both K-NN and Random Forest. The classification performance of K-NN and Random Forest is similar. Table IV shows the average classification results of the three classifiers, when the simulation times is set to 10. According to Table IV, the recognition rate of the PSO-SVM classifier reaches to $90\%$. The average recognition rate of Random Forest and K-NN is around $80\%$. Moreover, according to Tables I, II and III, the recognition rate of PSO-SVM achieves the highest accuracy for four emotions by comparing with the results of three classifiers.  

\section{Conclusion}

In this paper, we presented a feasible method of emotion recognition based on ECG signal. From analysis results of emotion data, we can see that the PSO-SVM scheme has the best classification performance for the ECG signals. The obtained results is better than previous researches in the emotion detection based on ECG signal. In addition, according to the classification consequences, the effectiveness of feature extraction was proved. The average detection rate experienced feature feature extraction has reached $90\%$ and results of previous research are around $80\%$. The recognition rate approached to the one based on facial recognition, but the method based on ECG signal is much simpler on steps and data processes. In addition, the ECG signals can truly react the emotion of individuals, because it is difficult for individuals to mask the change on ECG signals. In our future work, we will apply radio sensing techniques, such as \cite{IREALCARE5,GI1,GI2,GI3,GI4}, instead of wearable devices for emotion recognition. We will also develop privacy preservation algorithms \cite{privacy1,privacy2,privacy3,privacy4} to protect the users' privacy.

\end{document}